\documentstyle[12pt,boxedminipage,epsfig,wrapfig,floatflt,shadow]{article}

\begin{document}

\title{Centripetal Acceleration: Often Forgotten or Misinterpreted}

\author{Chandralekha Singh\\
 Department of Physics and Astronomy\\
 University of Pittsburgh, Pittsburgh, PA, 15260}

\date{ }

\maketitle
\begin{abstract}
Acceleration is a fundamental concept in physics which is taught in
mechanics at all levels. Here, we discuss some challenges in teaching
this concept effectively when the path along which the object is moving
has a curvature and centripetal acceleration is present. We discuss
examples illustrating that both physics teachers and students have
difficulty with this concept. We conclude with instructional strategies
that may help students with this challenging concept. 
\end{abstract}

\section{Introduction}

Helping students distinguish between velocity and acceleration has
always been challenging for physics instructors.~\cite{anti1,anti2} Several instructors
and researchers have analyzed the difficulties in teaching velocity and acceleration
to introductory physics students in different contexts.~\cite{lillian,acceleration,velocity,fred2,fred}
However, teaching the concept of acceleration effectively still remains elusive. Acceleration is the rate of
change of velocity with time. Since velocity is a vector, the velocity can change because its magnitude (speed) changes,
its direction changes, or both. For example, if a car moves in a straight line but slows down or speeds up, its only the speed that
is changing and not its direction. For uniform circular motion of an object, the speed is constant and it is only the
direction of velocity that is changing. In this case, the velocity is tangent to the path, and acceleration, called centripetal
acceleration, is towards the center of the circle at each instant. Since the speed is not changing, the velocity and centripetal acceleration
are always perpendicular to each other. In a more general motion of an object, both the magnitude and direction
of velocity could change and we can have two perpendicular non-zero components of acceleration: the tangential component $a_t$
due to the changes in the speed, and the centripetal acceleration $a_c$ due to the changes in the direction of velocity.
Here, we discuss situations involving centripetal acceleration which are universally challenging.

In a recent article~\cite{teacher},
the following two multiple-choice questions about the acceleration of a rolling ball on
a ramp of the shape shown in Figure 1 were given:
``A ball rolls on a ramp as shown. As it rolls from A to B its velocity increases and its acceleration:
(A) increases also, (B) decreases, (C) remains constant.
And when the ball rolls from B to C its acceleration:
(D) increases, (E) decreases, (F) remains constant."
The answers were given to be A and E respectively with the explanation
``Acceleration depends on the slope of the ramp. The slope
of the ramp gradually increases between points A and B, so acceleration
there gradually increases. But between points B and C, the slope gradually
decreases. This means acceleration beyond B gradually decreases also.\char`\"{}~\cite{teacher}
However, answer E for the second part is incorrect because it fails to account for centripetal
acceleration $a_c=v^2/R$, where $v$ is the speed and
$R$ is the radius of curvature.~\cite{lasota,hewitt2} The second question asks about the acceleration
as the particle rolls from point B to C and not about the magnitude
of the tangential acceleration $a_t$ alone. Therefore, the correct
answer should be {}``The magnitude of acceleration from point B to C may increase
or decrease depending upon the radius of curvature and the speed of
the object at point C\char`\"{}. 
It should be noted that point C is still part of the curved surface, i.e.,
at point C the ball is not rolling on a horizontal surface (infinite curvature).
As we shall see below, for the drawing
used to illustrate the situation in Figure 1, the magnitude of acceleration actually
increases from point B to point C.

\section{Calculation of the Magnitude of Acceleration}

Let's first calculate the magnitude of acceleration at point B based on the drawing in Figure 1. 
Figure 1 shows the estimates obtained for the three critical dimensions in the drawing: $H=8.6$ cm,
the height of the hill, $\theta=67^0$, the angle that the tangent to the path makes with the horizontal at the point of inflection
(B), and $R=6.1$ cm, the radius of curvature from point B to point C. 

First we focus on calculating the acceleration magnitude at point B. 
To simplify the calculation, we will assume that the ball is simply
sliding down the hill rather than rolling (later we will restore the
rolling motion). Since there is no curvature at the point of inflection,
the centripetal acceleration at point B is zero and the total acceleration
is only due to the tangential component. This acceleration has a magnitude
$a_t=gsin(\theta)$ where $g$ is the magnitude of the acceleration due to gravity.

Next, we assume that at point C, the slope of the curve is so small that the 
tangential acceleration can be ignored and the total acceleration is simply the centripetal acceleration.
The centripetal acceleration can be calculated as $a_{c}=v^{2}/R$, where we must still determine the speed of the ball $v$. We may
assume that the ball starts from rest at point A and ignore air resistance,
so that total mechanical energy is conserved along the way between points A and C. Hence,
$m g H = 1/2 m v^2$ and therefore, solving for centripetal acceleration $a_c=2 g H / R$. Note
that the answer only depends on the ratio H/R, so the answer is independent of the overall scale factor for the drawing. 
The ratio of the magnitudes of acceleration at points C and B is $a_c/a_t=2H/(R sin(\theta))$. 
Because the factor $2 H/R > 1$ for parameters in Figure 1, we see that regardless of the slope of the
incline at point B, the acceleration at point C must be larger than that at point B.

Now we include the effects due to rolling motion. As it turns out,
the rolling has no impact whatsoever on the above result. We assume that the ball has a moment of inertia 
$I=K{}mr^2$, where K=2/5 for a uniform solid sphere rolling about its center and $r$ is the radius of the ball.
Analysis of the force and torque equations
at point B yields a modified result: $a_t=gsin(\theta)/(1+K)$.~\cite{calc} 
A similar analysis at point C, taking into account both linear and rotational
kinetic energies in the conservation of energy equation and using $v=r \omega$ for rolling (here $\omega$ is the angular speed and the
rotational kinetic energy is $1/2 I \omega^2$), yields a similarly modified result: $a_{c}=v^{2}/R=2gH/(R(1+K))$.
Both factors of 1/(1+K) cancel if we take the ratio of the magnitudes of acceleration 
$a_c/a_t=2H/(R sin(\theta))$ at points C and B similar to the case when the
object was sliding rather than rolling. Thus, the total acceleration at point C will be larger
than that at point B if $2H/(R sin(\theta))>1$ 

\section{Centripetal Acceleration of a Pendulum}

Failure to include changes in direction of an object as being accelerations (as
opposed to just changes in speed) is widespread among novices and
experts alike. In a survey conducted by Reif et al.~\cite{fred} related
to acceleration at different points on the trajectory of a simple
pendulum, similar difficulties were found. Reif
et al. asked several Physics Professors at the University of California,
Berkeley who had taught introductory physics recently to explain how
the acceleration at various points on the trajectory of the simple
pendulum changed. A surprisingly large fraction of professors incorrectly
noted that the acceleration at the lowest point of the trajectory
is zero because they did not account for the centripetal acceleration.
When they were explicitly asked to reconsider their responses, approximately half
of them noticed that they were forgetting to take into account the
centripetal acceleration whereas the other half continued with their
initial assertion that the acceleration is zero at the lowest point
in the trajectory. In another study~\cite{edit}, many introductory
physics students had similar difficulties when calculating the normal
force on an object which is moving along a circular path and is at
the highest point.

\vspace*{-.1in}
\section{Acceleration is Difficult to Visualize but Easy to Feel}
\vspace*{-.1in}

There have been some interesting studies of students' perception of
motion~\cite{mestre} while watching objects moving on a computer
screen. However, further exploration is required to develop a more precise understanding
of human perception of velocity and acceleration. One difficulty is
that while humans can visually obtain a reasonably good sense of the speed of
an object quickly, the magnitude of acceleration of an object is difficult
to gauge visually. To illustrate, consider the following situation
that occurs commonly at a traffic intersection (Figure 2). Car
A and its passengers are travelling at a constant speed $V_A$. Car B and its passengers
are travelling with speed $V_B$ toward the intersection. The speed
of car B is such that if there is no deceleration, there will be a
collision between the two cars. This situation is ordinarily disturbing,
but only passengers in Car A are worried. The reason is that car B is decelerating.
Passengers in Car B can sense the deceleration and combine that information with
visual cues about the speed to help determine that a collision is
not imminent. Passengers in Car A receive no such inertial cues quickly, and must use
higher-level cognitive tools such as watching Car B over a period of time and calculating the
change in velocity per unit time implicitly in their minds to deduce that there will not be a collision.

The human body is an accelerometer and a person who is accelerating
has a {}``feel\char`\"{} for both the magnitude and direction of acceleration. For example, when our car
takes off from rest in a straight line, the acceleration is in the
same direction as the velocity but our body feels as if it is thrown backwards
in a direction opposite to the acceleration. Similarly, if the
car comes to rest, the acceleration is opposite to the direction of
velocity but the body lunges forward in a direction opposite to the
acceleration. These are demonstrations of inertia. Even a blind-folded person can tell that he/she is accelerating
when going up and down the curves of a roller coaster. When making
a turn in a circle, the acceleration is towards the center of the
circle (if the tangential component is zero) but our body feels as if it is thrown outward opposite to the direction
of the acceleration. If students are encouraged to imagine what they
will feel if they were experiencing a particular motion, they may
develop a better intuition for the concept of acceleration.

\vspace*{-.1in}
\section{Other Misconceptions about Centripetal Acceleration}
\vspace*{-.1in}

The mass of an object multiplied by its centripetal acceleration is often
called the {}``centripetal force\char`\"{}. This jargon is unfortunate
because one common misconception that students have about centripetal
force is that it is a new physical force of nature rather than a component of the net
force. For example, when we asked introductory physics students to
find the magnitude of the normal force (apparent weight) on a person moving
along a curved path at a time when the person is at the top of the curve,
many students incorrectly used {\bf equilibrium} application of Newton's Second Law.~\cite{edit}
Several students who were systematic in their problem solving, and drew free
body diagrams, wrote down the equation for the {\bf equilibrium} application of Newton's second law.
In their free body diagram, they incorrectly included centripetal force when passing over the highest point on the curve 
as if it was a separate physical force in addition to the normal force and the gravitational force. 
Using the equilibrium application of Newton's second law, students incorrectly
obtained that the magnitude of the normal force on the person
of mass $m$ moving with a speed $v$ along a circular curve of radius
$r$ is $F_{N}=mg+mv^{2}/r$ rather than the correct expression $F_{N}=mg-mv^{2}/r$.
Hence, they concluded that the apparent weight of the person increases
rather than decreases when passing over the bump. It is possible
that if students had invoked their intuition and imagined how they
will feel when moving over the circular bump, they may have realized
that their apparent weight cannot be larger than $mg$ when moving over the
bump. Another difficulty is that students often consider the pseudo forces, e.g., the centrifugal force, 
as though they were real forces acting in an inertial reference frame. It is advisable to avoid
discussions of pseudo forces altogether because they are likely to confuse students further.

\vspace*{-.1in}
\section{Summary and Instructional Implications}
\vspace*{-.1in}

We discussed examples to illustrate that centripetal acceleration
is a challenging concept even for physics teachers. Sometimes, veteran
physics teachers miss centripetal acceleration when calculating the
magnitude of total acceleration of an object moving along a curved
path. Students have additional difficulty with this concept and often
believe that mass times centripetal acceleration, termed centripetal
force, is a new physical force of nature rather than a net force.

Tutorials in introductory physics~\cite{tutor} have been found to
improve students' understanding of concepts related to acceleration.
Another instructional strategy that can help students with the concept
of acceleration is providing them with an opportunity for kinesthetic
explorations~\cite{pec}. Since human body can sense acceleration, such
explorations can help them understand and remember related concepts and reason intuitively about problems 
involving acceleration~\cite{intuition}.

\pagebreak

\begin{center}
\epsfig{file=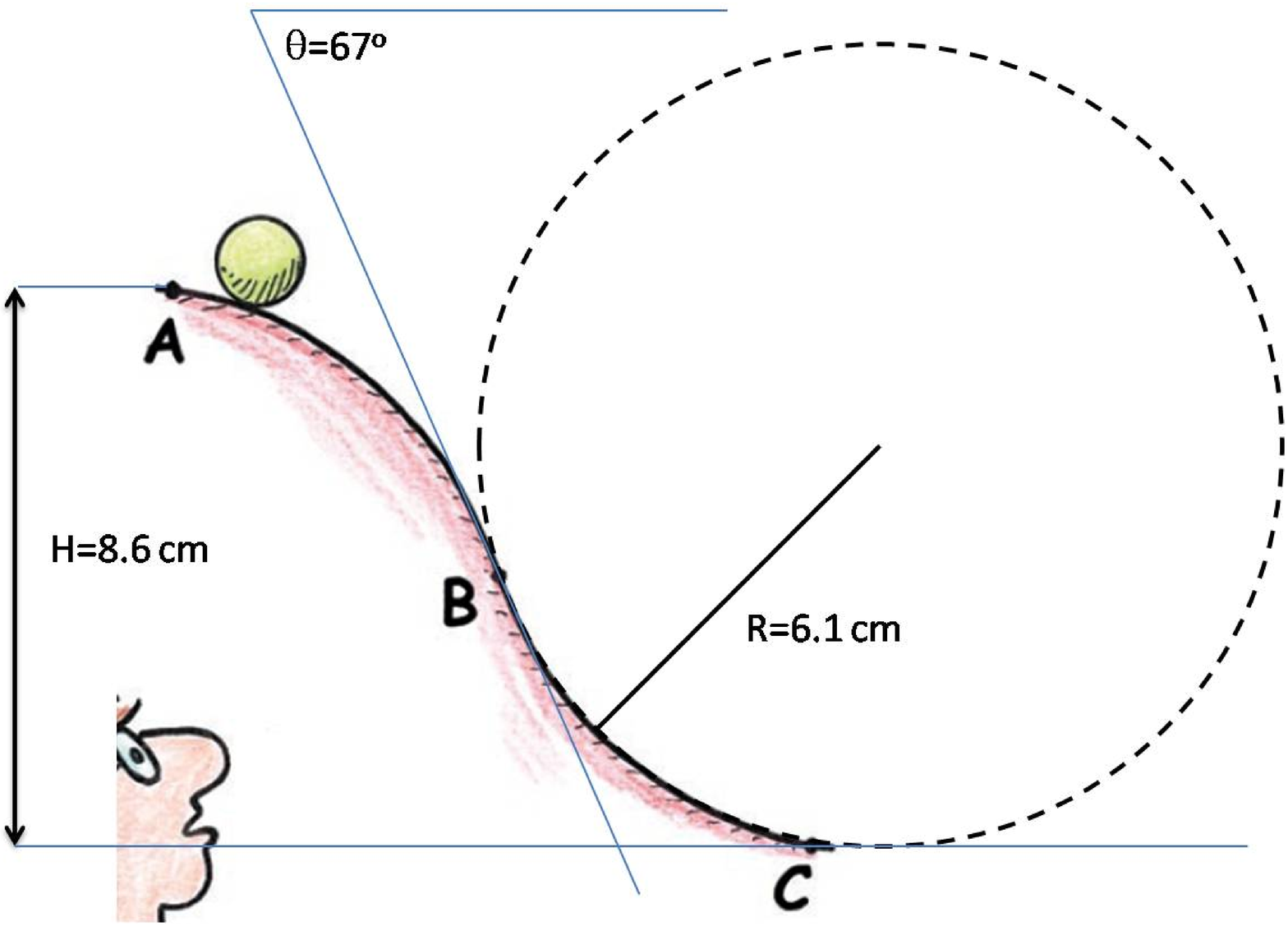,height=5.0in} 
\par\end{center}

Figure 1: Using geometry, the angle that the tangent to the path at
point B makes with the horizontal is $67^{0}$. The difference in height
between points A and C is $H=8.6$ cm and the radius $R=6.1$ cm. We note that it is
only the ratio $H/R$ that is important for determining the centripetal
acceleration at point C.

\pagebreak

\begin{center}
\epsfig{file=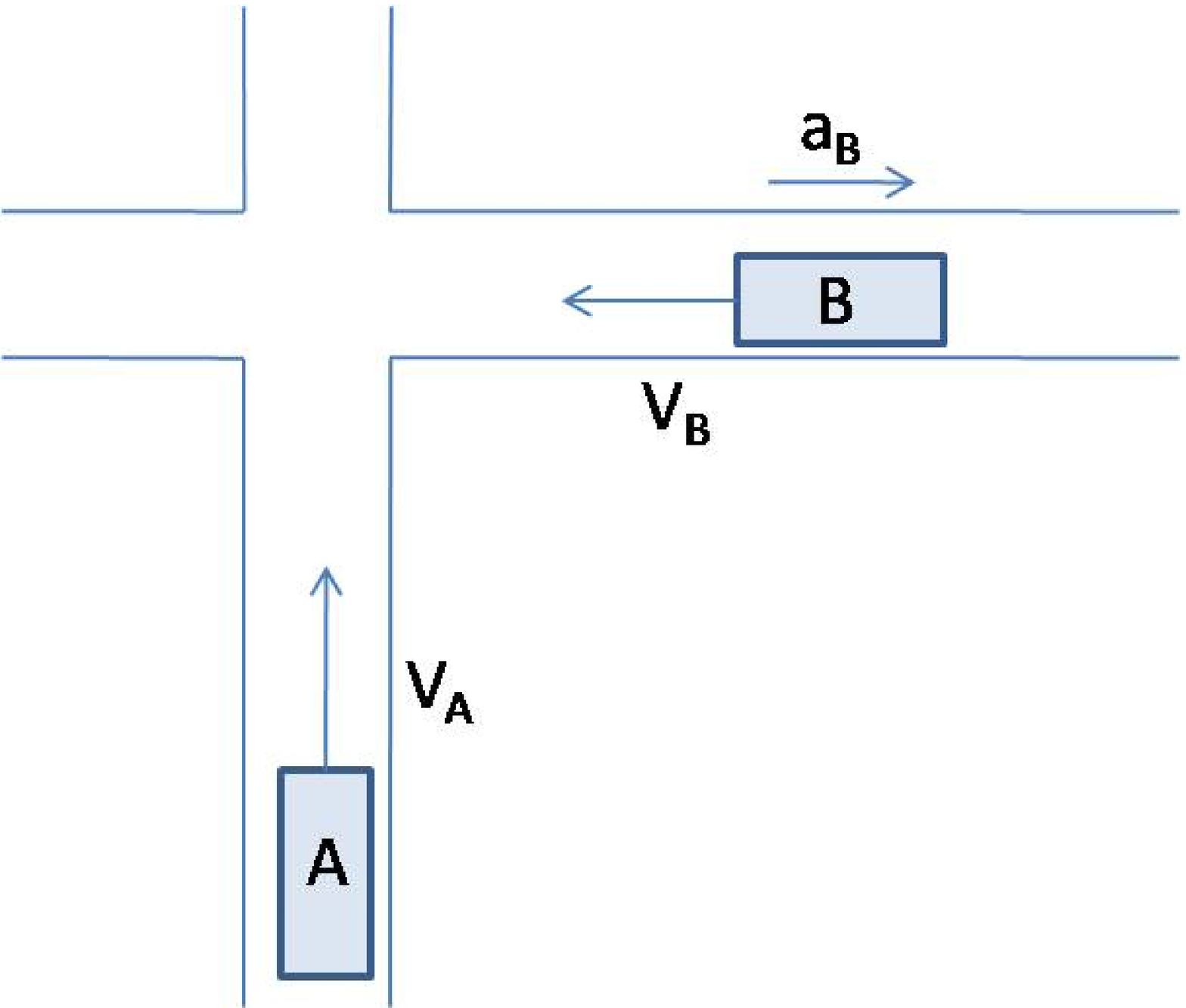,height=3.in} 
\par\end{center}

Figure 2: A diagram of the intersection with two cars. Car B has a large speed but is decelerating. The passengers in Car A
can quickly obtain a good feel for the speed of Car B but not as quickly for its acceleration (deceleration in this case).


\begin{thebibliography}{10}
\bibitem{calc} To derive the expression for the tangential acceleration $a_t$ at point B when the ball is rolling, 
we can draw a free body diagram on the incline with the gravitational force $mg$, normal force and the static frictional
force $f_s$ and use Newton's second law for linear motion to conclude that $mg sin(\theta)-f_s=ma_t$. Similarly,
using Newton's second law for rotational motion, we have net torque due to friction about the center of the ball 
$f_s r=I \alpha$. Using $I=K m r^2$ and $\alpha=a_t/r$ for rolling, we have $f_s r=K m r a_t$. Plugging the value of static
frictional force in $mg sin(\theta)-f_s=m\,\, a_t$, we obtain $a_t=g sin(\theta)/(1+K)$ as desired.
\end{thebibliography}
\end{document}